\documentstyle[preprint,prd,eqsecnum,aps,epsf]{revtex}

\tightenlines
\begin{document}

    \newcommand{\DSC}{D\hspace{-0.25cm}\slash_{\bot}}
    \newcommand{\DSP}{D\hspace{-0.25cm}\slash_{\|}}
    \newcommand{\DS}{D\hspace{-0.25cm}\slash}
    \newcommand{\DC}{D_{\bot}}
    \newcommand{\DSCX}{D\hspace{-0.20cm}\slash_{\bot}}
    \newcommand{\DSPX}{D\hspace{-0.20cm}\slash_{\|}}
    \newcommand{\DP}{D_{\|}}
    \newcommand{\QV}{Q_v^{+}}
    \newcommand{\QVB}{\bar{Q}_v^{+}}
    \newcommand{\QVP}{Q^{\prime +}_{v^{\prime}} }
    \newcommand{\QVBP}{\bar{Q}^{\prime +}_{v^{\prime}} }
    \newcommand{\QVHZ}{\hat{Q}^{+}_v}
    \newcommand{\QVHZB}{\bar{\hat{Q}}_v{\vspace{-0.3cm}\hspace{-0.2cm}{^{+}} } }
    \newcommand{\QVPHZB}{\bar{\hat{Q}}_{v^{\prime}}{\vspace{-0.3cm}\hspace{-0.2cm}{^{\prime +}}} }
    \newcommand{\QVPHFB}{\bar{\hat{Q}}_{v^{\prime}}{\vspace{-0.3cm}\hspace{-0.2cm}{^{\prime -}} } }
    \newcommand{\QVPHB}{\bar{\hat{Q}}_{v^{\prime}}{\vspace{-0.3cm}\hspace{-0.2cm}{^{\prime}} }   }
    \newcommand{\QVHF}{\hat{Q}^{-}_v}
    \newcommand{\QVHFB}{\bar{\hat{Q}}_v{\vspace{-0.3cm}\hspace{-0.2cm}{^{-}} }}
    \newcommand{\QVH}{\hat{Q}_v}
    \newcommand{\QVHB}{\bar{\hat{Q}}_v}
    \newcommand{\VS}{v\hspace{-0.2cm}\slash}
    \newcommand{\MQ}{m_{Q}}
    \newcommand{\MQP}{m_{Q^{\prime}}}
    \newcommand{\QVHPMB}{\bar{\hat{Q}}_v{\vspace{-0.3cm}\hspace{-0.2cm}{^{\pm}} }}
    \newcommand{\QVHMPB}{\bar{\hat{Q}}_v{\vspace{-0.3cm}\hspace{-0.2cm}{^{\mp}} }  }
    \newcommand{\QVHPM}{\hat{Q}^{\pm}_v}
    \newcommand{\QVHMP}{\hat{Q}^{\mp}_v}

\draft
\title{ $B\to \rho l \nu$ Decay and $|V_{ub}|$}
\author{ W.Y. Wang and Y.L. Wu }
\address{Institute of Theoretical Physics, Chinese Academy of Sciences\\
 Beijing 100080, China }
\maketitle

\begin{abstract}
$B\to\rho l \nu$ decay is analyzed in the effective theory of heavy quark with infinite mass
limit. The matrix element relevant to the heavy to light vector meson semileptonic 
decays is parametrized by a set of four heavy flavor-spin independent universal 
wave functions at the leading order of effective theory. The form factors are calculated 
at the leading $1/m_Q$ order using the light cone sum rule method in the framework of 
effective theory. $|V_{ub}|$ is then extracted via $B\to \rho l\nu$ decay mode. 
\end{abstract}

\pacs{PACS numbers: 13.20.-v, 11.30.Hv, 11.55.Hx, 12.15.Hh, 12.39.Hg\\
 Keywords: heavy to light vector transition, $|V_{ub}|$, effective
 theory, light cone sum rule}

\newpage

\section{Introduction}\label{int}

$B\to \pi (\rho) l\nu$ decays have been studied in the full QCD via sum rule 
approach \cite{var,ar,arsc,pvda,pvmisu,pvesrb}, or using appropriate models 
\cite{myc60,cs45,myc9594}. In Refs.\cite{gzmy} and \cite{hly}, 
the heavy to light pseudoscalar and 
vector meson decay matrix elements in the heavy quark effective theory have been 
formulated respectively up to the order of $1/m_Q$. 
In Ref.\cite{bpi}, the leading order wave functions of $B\to \pi l\nu$ decay have been calculated 
in the effective theory of heavy quark by using the light cone sum rule method, 
and the important CKM matrix element $|V_{ub}|$ has been extracted. 

This short paper is parallel to Ref.\cite{bpi} and focuses on the calculation of the 
leading order wave functions of $B\to \rho l\nu$ decay and the extraction of 
$|V_{ub}|$ by using the light cone sum rules in the effective theory 
of heavy quark. It is organized as follows:
In section 2, we present the analytic formulae for $B\to \rho l\nu$ decay in 
the effective theory of heavy quark. In section 3, the relevant light cone sum 
rules are derived within the framework of effective theory. In section 4 we present 
the numerical results including the extracted value of $|V_{ub}|$. 
The value of $|V_{ub}|$ is compared with that extracted via $B\to \pi l\nu$ 
decay. A short summary is drawed in the last section. 

 \section{Matrix Element}\label{formulation}

The $B\to\rho l\nu$ decay matrix element is 
generally written in terms of four form factors as
\begin{eqnarray}
\label{fdef}
&& <\rho(p,\epsilon^*)|\bar u\gamma^\mu (1-\gamma^5)  b|B(p+q)>=-i (m_B+m_\rho) A_1(q^2)
     \epsilon^{*\mu} +i \frac{A_2(q^2)}{m_B+m_\rho} (\epsilon^{*}\cdot (p+q) )
    (2p+q)^\mu \nonumber\\
&& \hspace{2cm}+i\frac{A_3(q^2)}{m_B+m_\rho} (\epsilon^* \cdot (p+q)) q^\mu
   - \frac{2 V(q^2)}{m_B+m_\rho} \epsilon^{\mu \alpha \beta \gamma} 
    \epsilon^*_\alpha (p+q)_\beta p_\gamma,
\end{eqnarray}
where $p$ and $\epsilon^*$ are the momentum and polarization vectors of the 
$\rho$ meson, and $q$ is the momentum carried by the lepton pair. 

In the effective theory of heavy quark, heavy quark expansion (HQE) can be 
applied to the matrix element \cite{gzmy,hly,bpi,wwy,ww,excit}. 
The normalization of the matrix elements in full QCD
and in the effective theory is \cite{bpi,wwy,ww,excit}
\begin{eqnarray}
\label{normalization}
\frac{1}{\sqrt{m_B} }<\rho(p,\epsilon^*)|\bar u \Gamma b|B>=\frac{1}{\sqrt{\bar\Lambda_B}}
\{  <\rho(p,\epsilon^*)|\bar{u} \Gamma \QV|B_v>+O(1/m_b) \},
\end{eqnarray}
where $\bar\Lambda_B=m_B-m_b$, and 
\begin{equation}
\bar\Lambda=\lim_{m_Q\to \infty} \bar\Lambda_B
\end{equation} 
is the heavy
flavor independent binding energy reflecting the effects of the light degrees
of freedom in the heavy hadron. $\QV$ is the effective heavy quark field.

Based on the heavy quark symmetry (HQS), one can parametrize the leading order matrix element in the 
effective theory of heavy quark on the rhs. of eq. (\ref{normalization}) 
as \cite{hly} 
\begin{eqnarray}
\label{omegadef}
<\rho(p,\epsilon^*)|\bar{u} \Gamma \QV|B_v>=-i \mbox{Tr}[\Omega(v,p)\Gamma {\cal M}_v]
\end{eqnarray}
with the decomposition of $\Omega(v,p)$:
\begin{eqnarray}
\label{ABdef}
 \Omega(v,p)&=&L_1(v\cdot p) {\epsilon\hspace{-0.2cm}\slash}^* 
   +L_2( v\cdot p) (v\cdot \epsilon^*) +[L_3(v\cdot p) 
   {\epsilon\hspace{-0.2cm}\slash}^* +L_4(v\cdot p) (v\cdot \epsilon^* )]
   {\hat{p}\hspace{-0.2cm}\slash},   \nonumber\\  
\hat{p}^\mu&=&\frac{p^\mu}{v\cdot p} \;\;.
\end{eqnarray}
${\cal M}_v$ is the well known spin wave function associated with the heavy meson 
states,
\begin{eqnarray}
 \label{eq:spinwave1}
 {\cal M}_v=\sqrt{\bar{\Lambda}} \frac{1+v\hspace{-0.2cm}\slash}{2} \left\{
\begin{array}{ll}
 -\gamma_{5} & \mbox{for pseudoscalar meson} \\
 \epsilon\hspace{-0.15cm}\slash & \mbox{for vector meson with polarization vector}\; \; \epsilon^\mu .
\end{array}
 \right.
\end{eqnarray}
$L_i(i=1,2,3,4)$ are the leading order wave functions in the effective
theory. Generally, they do not depend on the heavy quark mass but are functions of the variable 
$v\cdot p$ and the energy scale $\mu$ as well. 
For the sake of simplicity 
we do not write down their $\mu$ dependence explicitly until the numerical analysis in 
section \ref{sum}.

Eqs. (\ref{fdef})-(\ref{eq:spinwave1}) lead to 
\begin{eqnarray}
\label{AandL}
A_1(q^2)&=&\frac{2}{m_B+m_\rho} \sqrt{\frac{m_B \bar\Lambda}{\bar\Lambda_B}} 
    \{ L_1(v\cdot p)+L_3(v\cdot p) \}  +\cdots ;  \nonumber\\  
A_2(q^2)&=&2 (m_B+m_\rho) \sqrt{\frac{m_B \bar\Lambda}{\bar\Lambda_B}} 
    \{\frac{L_2(v\cdot p)}{2 m^2_B} +\frac{L_3(v\cdot p)
    -L_4(v\cdot p)}{2m_B (v\cdot p)} \}  +\cdots ;  \nonumber\\ 
A_3(q^2)&=&2 (m_B+m_\rho) \sqrt{\frac{m_B \bar\Lambda}{\bar\Lambda_B}} 
    \{ \frac{L_2(v\cdot p)}{2 m^2_B} -\frac{L_3(v\cdot p)
    -L_4(v\cdot p)}{2m_B (v\cdot p)} \}   + \cdots ; \nonumber\\  
V(q^2)&=&\sqrt{\frac{m_B \bar\Lambda}{\bar\Lambda_B}}
    \frac{m_B+m_\rho}{m_B (v\cdot p) } L_3(v\cdot p) +\cdots .
\end{eqnarray}
The dots denote higher order $1/m_Q$ contributions which
will not be taken into account in the following calculations. 

\section{Light Cone Sum Rule}\label{sumrule}

The form factors $A_1,A_2,A_3$ and $V$ ( and/or $L_i(i=1,2,3,4)$ )  
are difficult to be predicted straightforwardly from QCD calculation 
due to their nonperturbative nature. 
QCD sum rules, quark models and lattice simulations are the main 
approaches to evaluate the nonperturbative contributions. 
For heavy to light decays, the light cone sum rule approach is 
found to be more reliable than the sum rules based on short distance 
operator product expansion (OPE) and vacuum condensate input. 
In the light cone sum rule calculation, the relevant correlation  
functions are expanded near the light cone, and the nonperturbative 
contributions are introduced into the treatment through the light 
cone distribution functions of the mesons. 

We now consider the vacuum-$\rho$ correlation function
\begin{eqnarray}
\label{correlator}
F^\mu(p,q)=i\int d^4x e^{-ip_{\small B}\cdot x} <\rho(p,\epsilon^*)|
   T\{\bar{u}(0)\gamma^\mu (1-\gamma^5)  b(0),\bar{b}(x)
  i\gamma^5 d(x)\} |0>.
\end{eqnarray}
Here the B meson has momentum $P_B=p+q$, whereas $p$ and $q$ are momenta 
carried by the rho-meson and leptons. 
Phenomenologically, we insert a complete set of states with B meson quantum 
numbers in (\ref{correlator}),
\begin{eqnarray}
\label{cor}
F^\mu(p,q)_{phen}&=&\frac{<\rho(p,\epsilon^*)|\bar{u}\gamma^\mu (1-\gamma^5) b|B>
   <B|\bar{b}i\gamma^5 d|0>} { m^2_B-(p+q)^2 }  \nonumber\\ 
&+&\sum_H \frac{<\rho(p,\epsilon^*)|
   \bar{u}\gamma^\mu (1-\gamma^5)b|H><H|\bar{b}i\gamma^5 d|0>}
   { m^2_H-(p+q)^2 }.
\end{eqnarray}
Due to eqs. (\ref{normalization}), (\ref{omegadef}) and the B decay constant 
definition \cite{ww}
\begin{eqnarray}
\label{Fdef}
<0|\bar{q} \Gamma \QV |B_v >=\frac{F}{2} \mbox{Tr}[\Gamma {\cal M}_v],
\end{eqnarray}
at the leading order of HQE, $F^\mu(p,q)_{phen}$ contributes
\begin{eqnarray}
\label{phen}
&& \frac{m_B \bar\Lambda}{m_b \bar\Lambda_B} \frac{2F}{2\bar\Lambda_B-2v\cdot k}
   \{ (L_1+L_3) \epsilon^{*\mu} -L_2 v^\mu (\epsilon^* \cdot v) 
   -(L_3-L_4) p^\mu \frac{\epsilon^* \cdot v}{v\cdot p} \nonumber\\
&&\hspace{3cm} +i\frac{L_3}{v\cdot p} \epsilon^{\mu\nu\alpha\beta} \epsilon^*_\nu p_\alpha 
   v_\beta \}+\int^\infty_{s_0} ds \frac{\rho(v\cdot p,s)}{s-2v\cdot k}
   +subtractions,
\end{eqnarray}
with $k^\mu$ being the heavy hadron's residual momentum, $k^\mu=P^\mu_B-m_b v^\mu$.
The integral represents the higher resonance contributions.

The correlator can also be calculated in theory and written as
\begin{eqnarray}
\label{theo}
\int^{\infty}_{0} ds \frac{\rho(v\cdot p,s)_{theory}}{s-2v\cdot k}+subtractions.
\end{eqnarray}
Furthermore the quark-hadron duality assumes the equality of $\rho(v\cdot p,s)_{theory}$ 
and the physical spectral density $\rho(v\cdot p,s)$.  
As a result, equating (\ref{phen}) and (\ref{theo}) yields
\begin{eqnarray}
\label{phentheo}
&& \frac{m_B \bar\Lambda}{m_b \bar\Lambda_B} \frac{2F}{2\bar\Lambda_B-2v\cdot k}
   \{ (L_1+L_3) \epsilon^{* \mu} -L_2 {v^\mu} ({\epsilon^*} \cdot v )
   -(L_3-L_4) p^\mu \frac{\epsilon^* \cdot v}{v\cdot p}
   +i\frac{L_3}{v\cdot p} \epsilon^{\mu\nu\alpha\beta} \epsilon^*_\nu p_\alpha 
   v_\beta \}  \nonumber\\ 
&&\hspace{2cm} =\int^{s_0}_{0} ds \frac{\rho(v\cdot p,s)}{s-2v\cdot k}+subtractions.
\end{eqnarray}

We now calculate the correlator (\ref{correlator}) in the 
effective theory of heavy quark. 
When neglecting higher order $1/m_Q$ corrections eq. (\ref{correlator}) can be written as 
\begin{eqnarray}
\label{correlatorinHQET}
 F^\mu(p,q)=i\int d^4x e^{-ip_B\cdot x +im_bv\cdot x}
 <\rho(p,\epsilon^*)|T\{\bar{u}(0)\gamma^\mu (1-\gamma^5)\QV(0), \QVB(x)i\gamma^5 d(x)\}|0>.
\end{eqnarray}
This could then be expanded into a series in powers of the twist of light cone distribution functions.
The $\rho$ distribution functions are defined by the following matrix elements
\cite{ar,pvda,pvmisu,pvesrb}.
\begin{eqnarray}
\label{wfdef}
<\rho(p,\epsilon^*)|\bar{u}(0) \sigma_{\mu\nu} d(x)|0>&=&-i f^\bot_\rho 
  (\epsilon^*_\mu p_\nu - \epsilon^*_\nu p_\mu)  
\int^1_0 du e^{iup\cdot x} \phi_\bot (u)  ,\nonumber\\
<\rho(p,\epsilon^*)|\bar{u}(0) \gamma_\mu d(x)|0>&=& f_\rho m_\rho p_\mu 
  \frac{\epsilon^*\cdot x}{p\cdot x} \int^1_0 du e^{iup\cdot x} \phi_{\|}(u) \nonumber\\ 
  &+& f_\rho m_\rho (\epsilon^*_\mu-p_\mu \frac{\epsilon^* \cdot x}{p\cdot x}) 
  \int^1_0 du e^{iup\cdot x} g^{(v)}_\bot (u), \nonumber\\
<\rho(p,\epsilon^*)|\bar{u}(0) \gamma_\mu \gamma_5 d(x)|0>&=& 
  \frac{1}{4} f_\rho m_\rho \epsilon_{\mu\nu\alpha\beta} \epsilon^{*\nu} p^\alpha x^\beta
  \int^1_0 du e^{iup\cdot x} g^{(a)}_\bot (u).
\end{eqnarray}
$\phi_\bot$ and $\phi_{\|}$ are twist 2 distribution functions of transversely and 
longitudinally polarized $\rho$ mesons respectively while $g^{(v)}_\bot $ and 
$g^{(a)}_\bot $ are associated with both twist 2 and twist 3 operators. 
Higher twist components still lack for more precise analysis, and they are beyond 
our consideration in this paper. 

Contracting the effective heavy quark fields into the propagator 
$\frac{1+v\hspace{-0.13cm}\slash}{2} \int^{\infty}_{0} dt \delta(-x-vt) $, 
eq.(\ref{correlatorinHQET}) turns into 
\begin{eqnarray}
\label{corresult}
F^\mu(y,\omega)&=&\frac{i}{2} \int^{\infty}_{0} dt \int^{1}_{0} du e^{\frac{it\omega}{2}}
  e^{-iuyt} \{ f_\rho m_\rho p^\mu \frac{\epsilon^* \cdot v}{p\cdot v} \phi_{\|} (u)
  +f_\rho m_\rho (\epsilon^{*\mu}-p^\mu \frac{\epsilon \cdot v}{p\cdot v})
   g^{(v)}_\bot(u)\nonumber\\
 &+& \frac{t}{4} f_\rho m_\rho \epsilon^{\mu\nu\alpha\beta} \epsilon^*_\nu p_\alpha v_\beta 
  g^{(a)}_\bot(u) +f^\bot_\rho [ (v\cdot p) \epsilon^{*\mu} -(v\cdot \epsilon^*) p^\mu 
  +i \epsilon^{\mu\nu\alpha\beta} \epsilon^*_\nu p_\alpha v_\beta ] \phi_\bot(u) \} 
\end{eqnarray}
with $y\equiv v\cdot p$ and $\omega \equiv 2v\cdot k$.

Performing a wick rotation of the t axis and using the feature of Borel transformation:
\begin{eqnarray}
   \hat{B}^{(\omega)}_T e^{\lambda \omega}=\delta(\lambda-\frac{1}{T}),
\end{eqnarray}
we get from (\ref{corresult})
\begin{eqnarray}
\label{BTcor}
&&\hat{B}^{(\omega)}_T F^\mu(y,\omega) =\int^1_0 du e^{-\frac{2}{T}uy} 
   \{ \frac{\epsilon^* \cdot v}{v\cdot p} p^\mu [f_\rho m_\rho (\phi_{\|}(u)
   -g^{(v)}_\bot(u) )-f^\bot_\rho (v\cdot p) \phi_\bot(u) ] \nonumber\\
&& \hspace{0.5cm}+ \epsilon^{*\mu} [f_\rho m_\rho g^{(v)}_\bot(u)+f^\bot_\rho (v\cdot p)
   \phi_\bot(u) ]+\epsilon^{\mu\nu\alpha\beta} \epsilon^*_\nu p_\alpha v_\beta
   [ \frac{-i}{2T} f_\rho m_\rho g^{(a)}_\bot(u)+if^\bot_\rho \phi_\bot(u) ]\}.
\end{eqnarray}

Following the approach in \cite{nr,pv}, we can now carry out continuous double Borel 
transformations on the correlator 
itself to produce the spectral function $\rho(y,s)$. 
\begin{eqnarray}
\label{spectralfun}
&&  \rho(y,s)=\hat{B}^{(-1/T)}_{1/s} \hat{B}^{(\omega)}_T F^\mu(y,\omega) \nonumber\\ 
&&\hspace{1.5cm} =\frac{1}{2y} \{ p^\mu \frac{\epsilon^* \cdot v}{v\cdot p}
    [f_\rho m_\rho (\phi_{\|}(u)-g^{(v)}_\bot(u))-f^\bot_\rho (v\cdot p) 
    \phi_\bot(u) ]+\epsilon^{*\mu} [f_\rho m_\rho g^{(v)}_\bot (u) \nonumber\\ 
&&\hspace{1.5cm} +f^\bot_\rho (v\cdot p) \phi_\bot(u) ]  
   +\epsilon^{\mu\nu\alpha\beta} \epsilon^*_\nu p_\alpha v_\beta 
   [ \frac{-i}{4y}f_\rho m_\rho (\frac{\partial}{\partial u } g^{(a)}_\bot(u) )
   +if^\bot_\rho \phi_\bot(u) ] \}_{u=\frac{s}{2y}}.
\end{eqnarray}
Eq.(\ref{spectralfun}) can be easily derived from (\ref{BTcor}) by first writing 
$\frac{1}{T}$ as a derivative of the exponent in (\ref{BTcor}) over $u$, and then 
using the method of integration by parts over $u$.

We then get from eqs.(\ref{phentheo}) and (\ref{spectralfun}) 
\begin{eqnarray}
\label{sr}
L_1(y)&=& \frac{m_b \bar\Lambda_B}{m_B \bar\Lambda} \frac{1}{4F} e^{2\bar\Lambda_B/T}
   \int^{s_0}_0 ds e^{-s/T} \frac{1}{y} f_\rho m_\rho [ g^{(v)}_\bot(u)+\frac{1}{4}
   (\frac{\partial}{\partial u} g^{(a)}_\bot (u) ) ]_{u=\frac{s}{2y}}, \nonumber\\
L_3(y)&=& \frac{m_b \bar\Lambda_B}{m_B \bar\Lambda} \frac{1}{4F} e^{2\bar\Lambda_B/T}
   \int^{s_0}_0 ds e^{-s/T} [\frac{-1}{4y} f_\rho m_\rho (\frac{\partial}{\partial u}
   g^{(a)}_\bot(u) )+f^\bot_\rho \phi_\bot(u) ]_{u=\frac{s}{2y}},\nonumber \\
L_4(y)&=& \frac{m_b \bar\Lambda_B}{m_B \bar\Lambda} \frac{1}{4F} e^{2\bar\Lambda_B/T}
   \int^{s_0}_0 ds e^{-s/T} \frac{1}{y} f_\rho m_\rho [ \phi_{\|}(u)-g^{(v)}_\bot(u)
   -\frac{1}{4} (\frac{\partial}{\partial u} g^{(a)}_\bot (u) ) ]_{u=\frac{s}{2y}},
\end{eqnarray}
and $L_2(y)$ equals zero in the present approximation since no twist 2 distribution 
functions contribute to it. For this reason, the form factors $A_2$ and $A_3$ have the 
same absolute value but opposite signs at the order considered in this paper.  

\section{Numerical Results}\label{sum}

The functions $\phi_{\bot}$ and $\phi_{\|}$
give the leading twist distributions in the fraction of total monmentum 
carried by the quark in transversely and longitudinally polarized mesons, 
respectively. 
They have a non-trivial scale dependence which 
can be described by the renormalization group method \cite{pvda}. 
These distribution functions can be expanded in Gegenbauer polynomials 
$C^{3/2}_n(x)$ whose coefficients are renormalized multiplicatively. 
Namely, writting their scale dependence explicitly, we have 
\begin{eqnarray}
\label{rhowfmu}
\phi_{\bot(\|)}(u,\mu)&=&6u(1-u)[ 1+\sum_{n=2,4,\cdots} a^{\bot(\|)}_n(\mu) 
   C^{3/2}_n(2u-1)  ], \nonumber\\
a^{\bot(\|)}_n(\mu)&=&a^{\bot(\|)}_n(\mu_0) (\frac{\alpha_s(\mu)}
   {\alpha_s(\mu_0)})^{(\gamma^{\bot (\|)}_n-\gamma^{\bot (\|)}_0)/(2\beta_0)},
\end{eqnarray}
where $\beta_0=11-(2/3)n_f$, and the one loop anomalous dimensions are \cite{dfmn,mm} 
\begin{eqnarray}
\label{anoma}
\gamma^{\|}_n&=&\frac{8}{3}(1-\frac{2}{(n+1)(n+2)}+4\sum^{n+1}_{j=2} \frac{1}{j}),\nonumber\\
\gamma^{\bot}_n&=&\frac{8}{3}(1+4\sum^{n+1}_{j=2} \frac{1}{j}).
\end{eqnarray}
The coefficients $a^\bot_n$ and $a^{\|}_n$ themselves are nonperturbative 
parameters, and have been calculated using sum rule methods in Ref.\cite{pvda}. 
In the following discussions we will use the values \cite{pvda}
\begin{eqnarray}
\label{para}
a^\bot_2(1\mbox{GeV})=0.2\pm 0.1, \;\;\; a^{\|}_2(1\mbox{GeV})=0.18\pm 0.10 
\end{eqnarray}
and $a^{\bot(\|)}_n=0$ for $n\neq 2$.

The functions $g^{(v)}_\bot$ and $g^{(a)}_\bot$ describe 
transverse polarizations of quarks in the longitudinally polarized 
mesons. They receive contribtuions of both twist 2 and twist 3. 
And the twist 2 contributions are related to the longitudinal
distribtuion $\phi_{\|}(u,\mu)$ by Wandzura-Wilczek type relations
\cite{pvda,pvmisu}:
\begin{eqnarray}
\label{gcva}
g^{(v),twist \;\; 2}_\bot (u,\mu)&=&\frac{1}{2}[\int^u_0 dv \frac{\phi_{\|}(v,\mu)}
  {1-v} +\int^1_u dv \frac{\phi_{\|}(v,\mu)}{v} ], \nonumber\\
g^{(a),twist \;\; 2}_\bot (u,\mu)&=&2[(1-u)\int^u_0 dv \frac{\phi_{\|}(v,\mu)}
  {1-v} +u \int^1_u dv \frac{\phi_{\|}(v,\mu)}{v} ].
\end{eqnarray}

For the energy scale $\mu$ to be used in the sum rules (\ref{sr}), we use 
\begin{equation}
\label{mub}
\mu_b \sim \sqrt{m^2_B-m^2_b} \approx 2.4\mbox{GeV},
\end{equation}
which is an appropriate choice of scale set by the typical virtuality of 
the beautiful quark \cite{ar}.

The values of the hadron quantities 
$f_\rho$, $f^\bot_\rho$, $\bar\Lambda_B$, $\bar\Lambda$ and $F$ are needed 
in order to perform the sum rule numerical analysis. 
The decay constant $f_\rho$ has been measured in experiments \cite{mab,lm}. 
$f^\bot_\rho$ is the tensor coupling defined by 
\begin{eqnarray}
<0|\bar{u} \sigma_{\mu\nu} d|\rho^+(p,\epsilon)>=i(\epsilon_\mu p_\nu
  -\epsilon_\nu p_\nu ) f^\bot_\rho
\end{eqnarray}
and its value has been calculated in Ref.\cite{pvda}.
$\bar\Lambda_B$, $\bar\Lambda$ and $F$ are associated with heavy 
mesons and are parameters in the effective theory of heavy quark. 
Their values have been estimated consistently in Ref.\cite{ww} by sum rules in the framework of 
effective theory. We use for these parameters the following values, 
\begin{eqnarray}
\label{paravalue}
&& f_{\rho^{\pm}}=(195\pm 7)\mbox{MeV}, \;\;\; 
f_{\rho^0}=(216\pm 5)\mbox{MeV}, \;\;\; 
f^\bot_\rho=(160 \pm 10 )\mbox{MeV}, \nonumber \\
&& \bar\Lambda_B \approx \bar\Lambda=0.53 \mbox{GeV}, \;\;\;\;
F=(0.30 \pm 0.06)\mbox{GeV}^{3/2}. 
\end{eqnarray}

Combining (\ref{AandL}) and (\ref{sr}), the form factors $A_1$, $A_2$, 
$A_3$ and $V$ can be calculated as functions of $T$, $q^2$ and $s_0$.
In Fig.1, we present our results for the form factors at 
the zero momentum transfer point $q^2=0$, 
which shows the variations of these form factors with respect to the 
Borel parameter $T$ at different threshold energy $s_0$. 
The $T$ range of interest should be similar to that in the light cone 
sum rule analysis for the $B\to \pi l\nu $ decay \cite{bpi}, i.e., $T\approx 2.0$GeV. 
It can be seen from Fig.1 that, in general consideration, the good stability 
of the form factors exists 
at the threshold $s_0=2.1 \pm 0.6$GeV.
With such a threshold energy, we are now in the stage to evaluate the 
$q^2$ dependence of the form factors. 
It should be intuitive and convenient to represent these form factors in an 
algebraic representation. We parametrize each form factor in terms of a 
set of three parameters as follows,
\begin{eqnarray}
\label{fitform}
F(q^2)=\frac{F(0)}{1-a_F q^2/m^2_B+b_F (q^2/m^2_B)^2},
\end{eqnarray}
where $F(q^2)$ can be any one of $A_1(q^2)$, $A_2(q^2)$, $A_3(q^2)$ and $V(q^2)$. 
The parameters $F(0)$, $a_F$ and $b_F$ can be fitted from the the sum rule 
results (\ref{sr}). The results at $s_0=2.1$GeV are presented in table 1.
The form factors as functions of the momentum transfer $q^2$ are also shown 
in Fig.2. 

\vspace{1.2cm}

\begin{center}
\begin{tabular}{c|c|c|c}
\hline \hline
\rule{0cm}{0cm} \hspace*{3cm}& \ \hspace*{3cm}& \hspace*{3cm}& \hspace*{3cm} \ \\ 
    & F(0) &   $a_F$ &  $b_F$   \\
\hline
$A_1$ & 0.257 &  0.352 & -0.239 \\
\hline
$A_2$ & 0.253   & 1.090 & 0.202   \\
\hline
$A_3$ & -0.253 &  1.090 & 0.202 \\
\hline
$V$ &  0.134  &  1.027  & -0.223 \\
\hline  \hline
\end{tabular}
\end{center}

\vspace{0cm}
\centerline{
\parbox{12cm}{
\small
\baselineskip=1.0pt
Table 1. Results of the three parameter fit (\ref{fitform}) for the $B\to \rho l\nu$ decay 
form factors. These data are fitted from the sum rules (\ref{sr}) at the Borel 
parameter $T=2.0$ GeV and the threshold energy $s_0=2.1 $ GeV.
} }

\vspace{1.2cm}

When the lepton masses are neglected, the differential decay width of 
$B\to \rho l\nu $ with respect to the momentum transfer $q^2$ is \cite{pvmisu} 
\begin{eqnarray}
\label{gammaq2}
\frac{d\Gamma}{dq^2}=\frac{G^2_F |V_{ub}|^2}{192 \pi^3 m^3_B} 
  \lambda^{1/2} q^2 (H^2_0+H^2_+ + H^2_-) 
\end{eqnarray}
with the helicity amplitudes 
\begin{eqnarray}
\label{heliampli}
H_{\pm}&=&(m_B+m_\rho) A_1(q^2) \mp \frac{\lambda^{1/2}}{m_B+m_\rho} V(q^2),\nonumber\\
H_0&=&\frac{1}{2m_\rho \sqrt{q^2}} \{ (m^2_B-m^2_\rho-q^2) (m_B+m_\rho) A_1(q^2) 
    -\frac{\lambda}{m_B+m_\rho}A_2(q^2)  \}
\end{eqnarray} 
and 
\begin{eqnarray}
\label{ladd}
\lambda \equiv (m^2_B+m^2_\rho-q^2)^2-4m^2_B m^2_\rho.
\end{eqnarray}
So, with the meson masses $m_B=5.28$GeV, $m_{\rho}=0.77$GeV and the maximum momentum 
transfer $q^2_{max}=m^2_B+m^2_\rho-2m_B m_\rho$, the integrated width 
of $B \to \rho l \nu$ turns out to be 
\begin{eqnarray}
\label{width}
\Gamma(B \to \rho l \nu)=(10.6 \pm 4.0 ) |V_{ub}|^2 \mbox{ps}^{-1}.
\end{eqnarray}
The error in eq.(\ref{width}) results from the variation of the threshold energy 
$s_0=1.5-2.7$GeV.

On the other hand, the branching fraction of $B^0 \to \rho^- l^+ \nu$ is 
measured by CLEO Collaboration \cite{cleo2},
$\mbox{Br}(B^0 \to \rho^- l^+ \nu)=(2.57\pm 0.29^{+0.33}_{-0.46} 
\pm 0.41) \times 10^{-4}$. With the 
world average of the $\mbox{B}^0$ lifetime \cite{pdg},
$\tau_{\tiny{\mbox{B}^0}}=1.56\pm 0.06 \;\mbox{ps}$, one has 
\begin{eqnarray}
\label{CLEO}
\Gamma(B^0\to \rho^- l^+ \nu)=(1.65 \pm 0.80 )\times 10^{-4} \mbox{ps}^{-1}.
\end{eqnarray}
From eqs.(\ref{width}) and (\ref{CLEO}) we get  
\begin{eqnarray}
|V_{ub}|=(3.9 \pm 0.6 \pm 0.5)\times 10^{-3},
\end{eqnarray}
where the first (second) error corresponds to the experimental (theoretical)
 uncertainty. Here the theoretical uncertainty is mainly
 considered from the threshold effects. In general, both the 
higher twist distribution functions and the QCD radiative corrections may 
modify the parameters and the sum rule results in eq.(\ref{sr}). 
The modification from higher distribution functions is hard to 
calculate at present because little is known about the high distribution 
functions themselves. It is found from the sum rule analysis that 
the two-loop QCD perturbative corrections may enlarge the constant 
$F$ by about 25\%, and increase $\bar\Lambda$ at the same time \cite{ww}. 
To be consistent, one should also include the QCD corrections for the correlation 
function at the same order. As such higher order QCD corrections to the 
correlator have not been considered here, we should take the two-loop 
QCD corrections of the decay constant $F$ as the theoretical uncertainty. 
Including this uncertainty, our final result for $|V_{ub}|$
is 
\begin{eqnarray}
\label{vub}
|V_{ub}|=(3.9\pm 0.6 \pm 0.7 )\times 10^{-3}.
\end{eqnarray}
This value may be compared with the one obtained in Ref.\cite{bpi} from 
$B\to \pi l\nu$ decay by using the same approach, 
\begin{equation}
\label{vubfrompi}
|V_{ub}|=( 3.4 \pm 0.5 \pm 0.5 )\times 10^{-3}.
\end{equation}
The coincidence of (\ref{vub}) and (\ref{vubfrompi}) within their errors 
proves the consistency of our light cone sum rule calculations 
of heavy to light semileptonic B decays in the framework of 
the effective theory of heavy quark. We also notice that the theoretical error 
in (\ref{vub}) is larger than that in (\ref{vubfrompi}), which is not out of  
expectation since we only include the leading twist distribution functions
in the $B\to \rho l\nu $ calculation.
This reflects the importance of a more complete study on the 
$\rho$ distribution functions. 

The estimate in eq.(\ref{vub}) is in agreement with that derived from full QCD
calculation \cite{ar}, and it is also close to the 
combined result from the analyses based on different models and treatments 
on $B\to \pi (\rho) l \nu$ transitions, 
\begin{equation}
\label{CLEOcomb}
|V_{ub}|=(3.25\pm 0.14^{+0.21}_{-0.29} \pm 0.55)\times 10^{-3},
\end{equation}
which is given by CLEO \cite{cleo2}.  

\section{Summary}\label{end}

We have studied $B\to \rho l\nu$ decay within the framework of effective
theory of heavy quark. 
In the effective theory, the relevant matrix elements can 
be expanded in powers of the inverse of the heavy quark mass. 
At the leading order approximation, the form factors 
concerned in this decay are related to four universal wave functions, 
which are independent of the heavy quark 
mass $m_Q$. 
Though the HQS loses some predictive power in the heavy to light 
decays, it would be helpful for relating different heavy to light 
decay channels. For example, $B\to \rho l\nu$ and 
$D \to \rho l\nu $ decays are characterized at the leading order 
of $1/m_Q$ by the same set of wave functions, $L_i(i=1,2,3,4)$.  

The form factors for $B\to \rho l\nu$ have been calculated in the effective theory 
using the light cone sum rule approach. 
The important CKM matrix element $|V_{ub}|$ has been extracted 
by comparing the values of integrated width obtained from sum rule 
calculations and from the experimental measurements.  
The result is
\begin{eqnarray}
|V_{ub}|=( 3.9 \pm 0.6  \pm 0.7 )\times 10^{-3}.
\end{eqnarray}
This result agrees with both the values  
extracted from the full QCD calculations and that from the 
$B \to \pi l\nu$ decay by using the same approach within the 
framework of the effective theory of heavy quark. 
This calculation has further shown the  
reliability of the heavy quark expansion and the predictive power of light
cone sum rule approach in studying heavy to light exclusive
decays. 
We have used $B\to \rho l\nu$ decay as an example for concret discussion. 
However, the method is general and the main formulae in this paper can be 
used to other heavy to light vector semileptonic decays after trival modifications 
such as simple replacement of some parameters.

In this paper, we have considered only the leading twist 2 distribution 
functions of $\rho$ meson. 
Both the higher twist and loop corrections and higher order $1/m_Q$ 
contributions should be included for a more accurate estimation of $|V_{ub}|$. 
It is noted that higher order $1/m_Q$ corrections may have different forms 
in the usual heavy quark effective theory and the new framework of heavy quark 
effective field theory \cite{wwy,yww,wy,ww,excit} due to the antiquark 
contributions.

\acknowledgments

This work was supported in part by the NSF of China under the
grant No. 19625514  as well as Chinese Academy of Sciences.


\newpage
\centerline{\large{FIGURES}}

\newcommand{\PICL}[2]
{
\begin{center}
\begin{picture}(500,170)(0,0)
\put(0,5){
\epsfxsize=7cm
\epsfysize=7cm
\epsffile{#1} }
\put(115,60){\makebox(0,0){#2}}
\end{picture}
\end{center}
}

\newcommand{\PICR}[2]
{
\begin{center}
\begin{picture}(300,0)(0,0)
\put(160,34){
\epsfxsize=7cm
\epsfysize=7cm
\epsffile{#1} }
\put(275,89){\makebox(0,0){#2}}
\end{picture}
\end{center}
}

\small
\mbox{}
{\vspace{1.2cm}}

\PICL{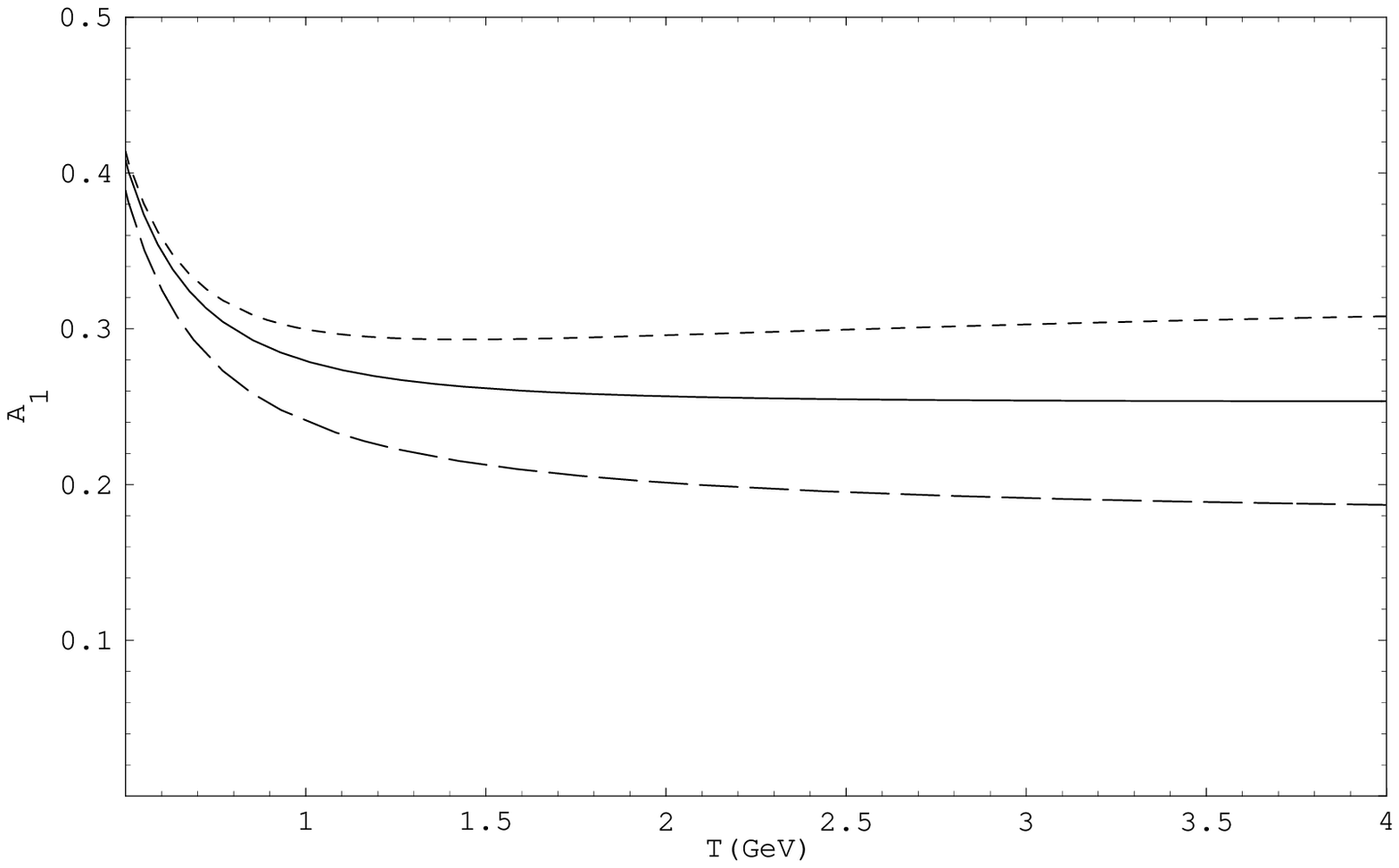}{Fig.1(a)}

\PICR{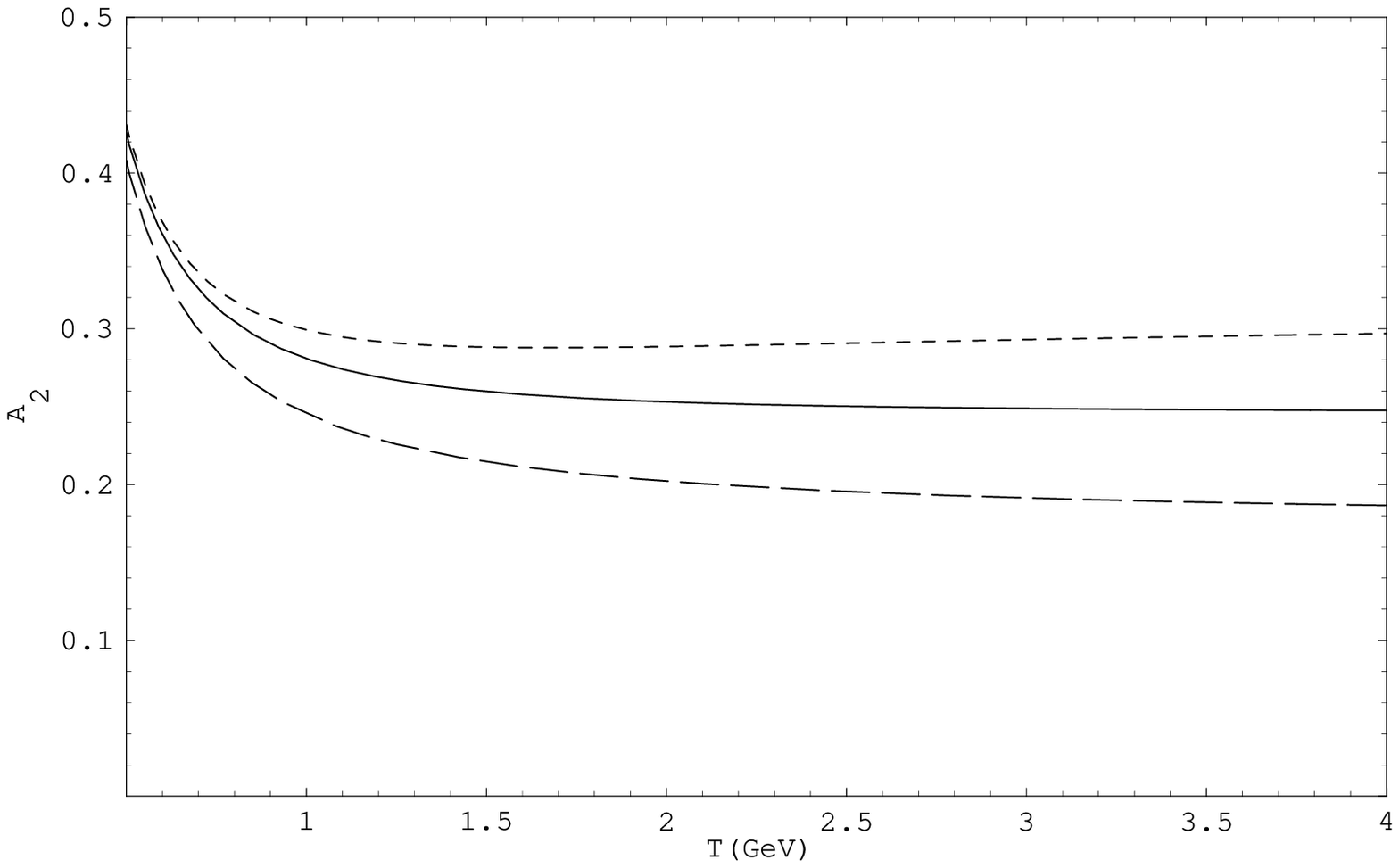}{Fig.1(b)}

\PICL{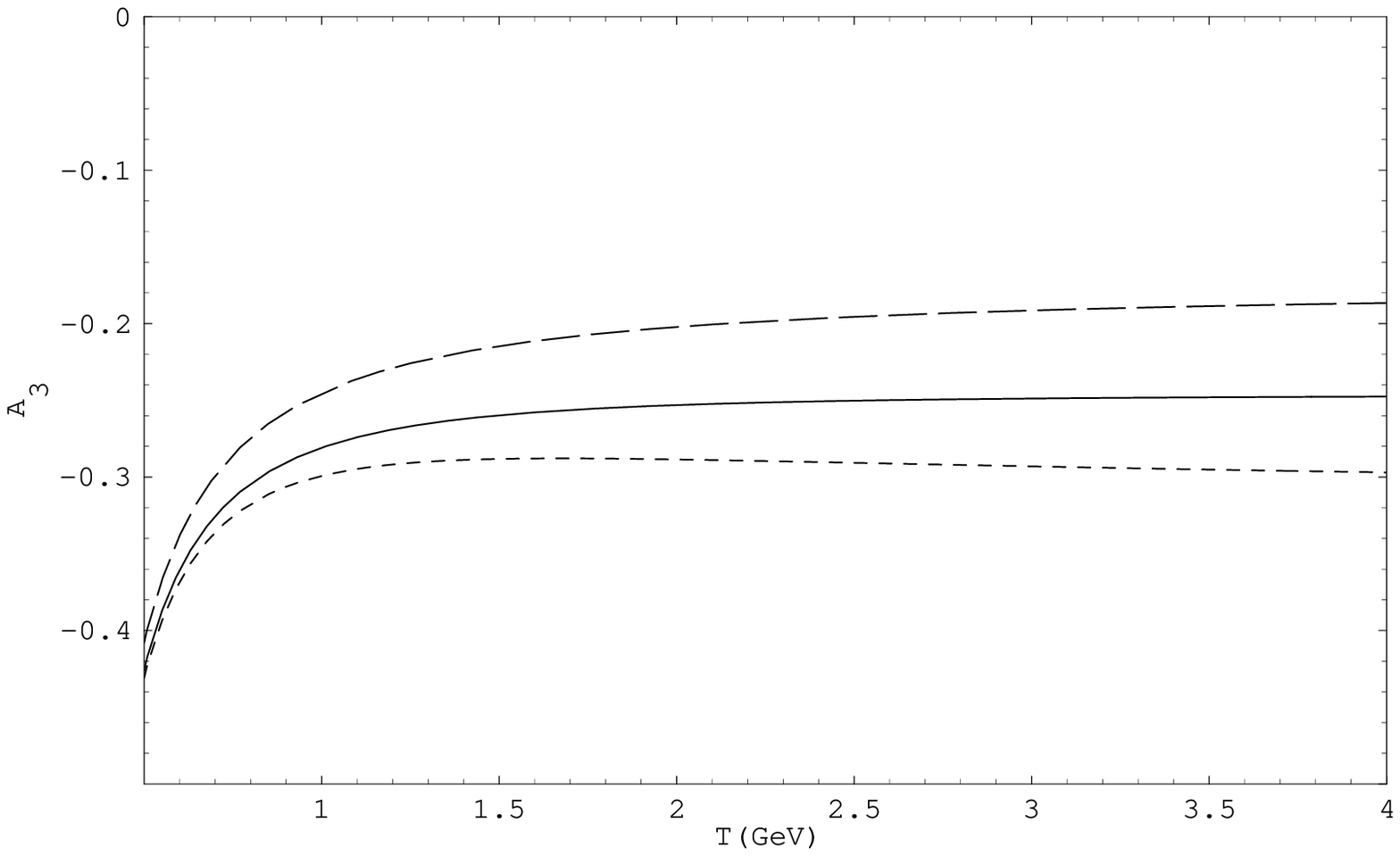}{Fig.1(c)}

\PICR{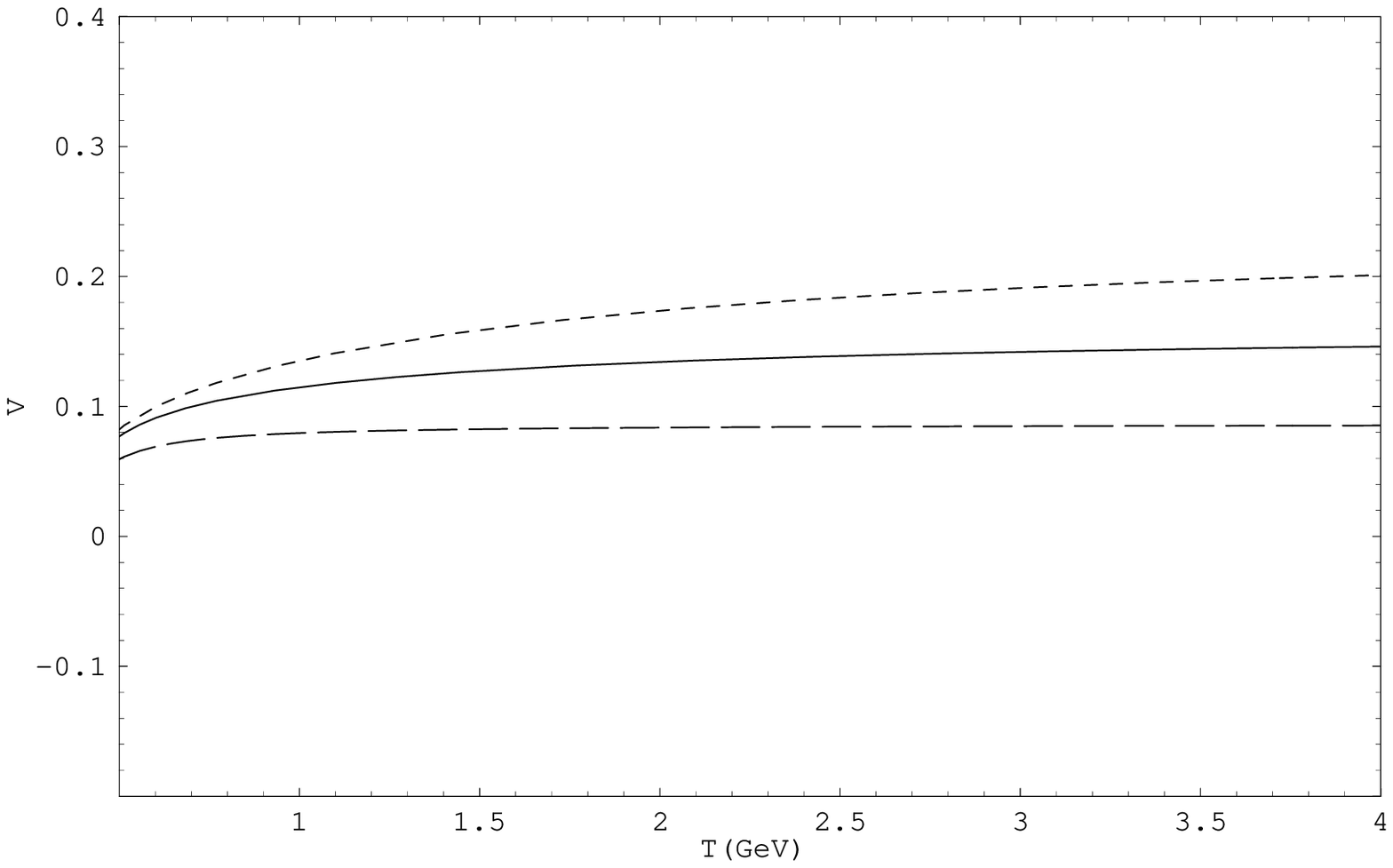}{Fig.1(d)}

\vspace{-2cm}
\centerline{
\parbox{12cm}{
\small
\baselineskip=1.0pt
Fig.1(a-d). Variation of form factors with the Borel parameter T
for different values of the continuum threshold $s_0$. The dashed, solid and
dotted curves correspond to $s_0=$1.5, 2.1 and 2.7 GeV respectively. 
Considered here is at the momentum transfer $q^2=0\mbox{GeV}^2$.}}

\newpage
\mbox{}
{\vspace{2cm}}

\PICL{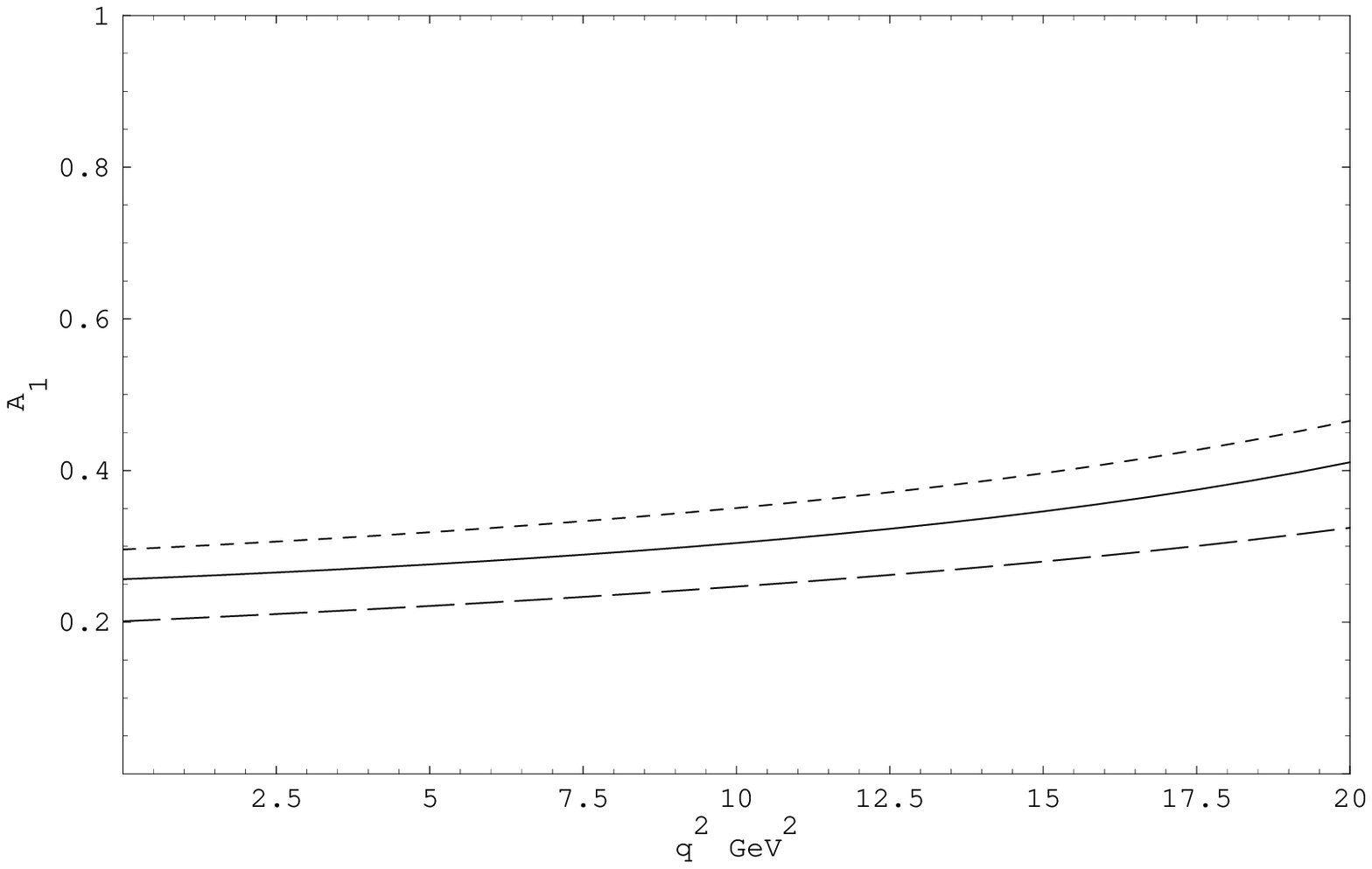}{Fig.2(a)}

\PICR{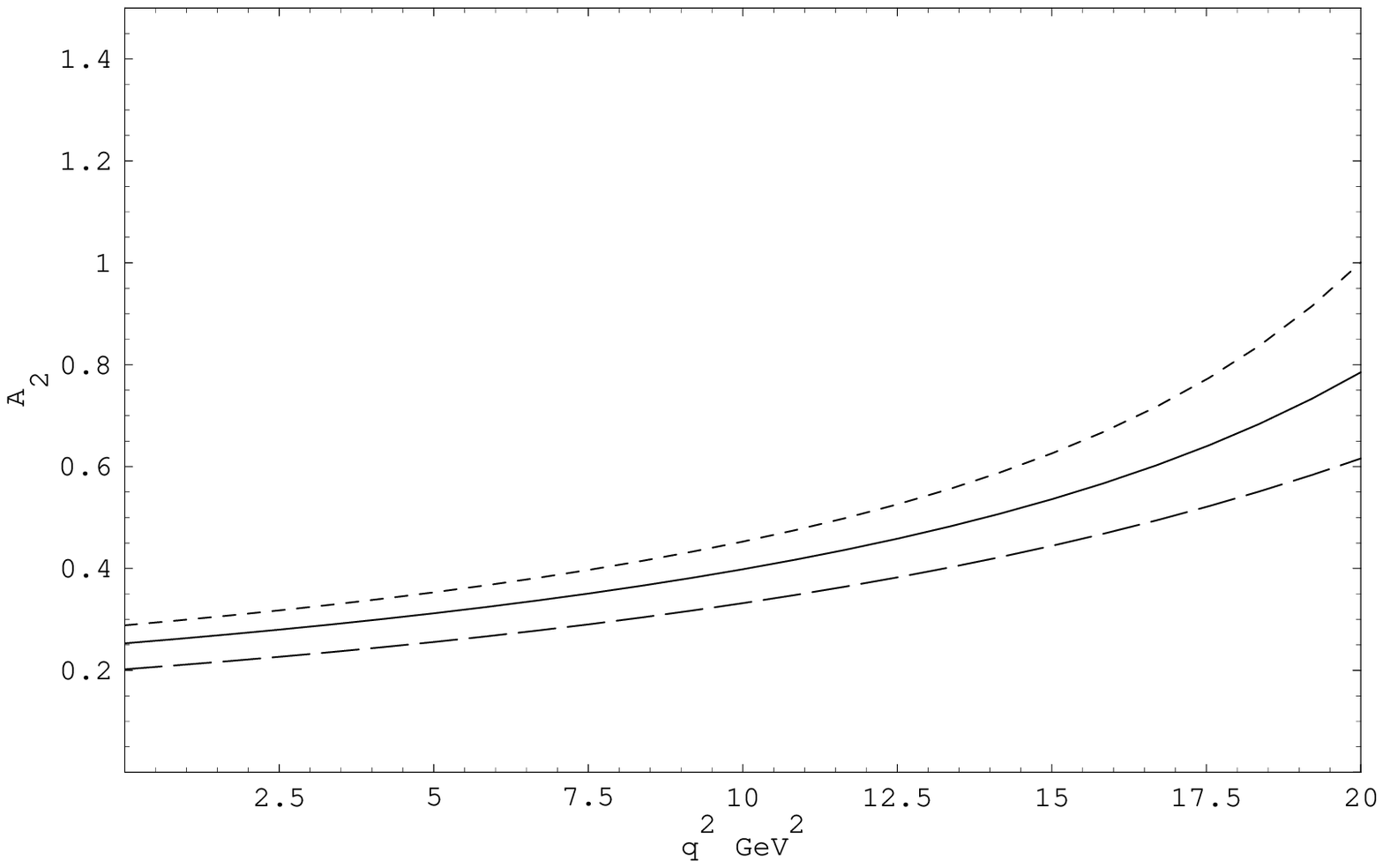}{Fig.2(b)}

\PICL{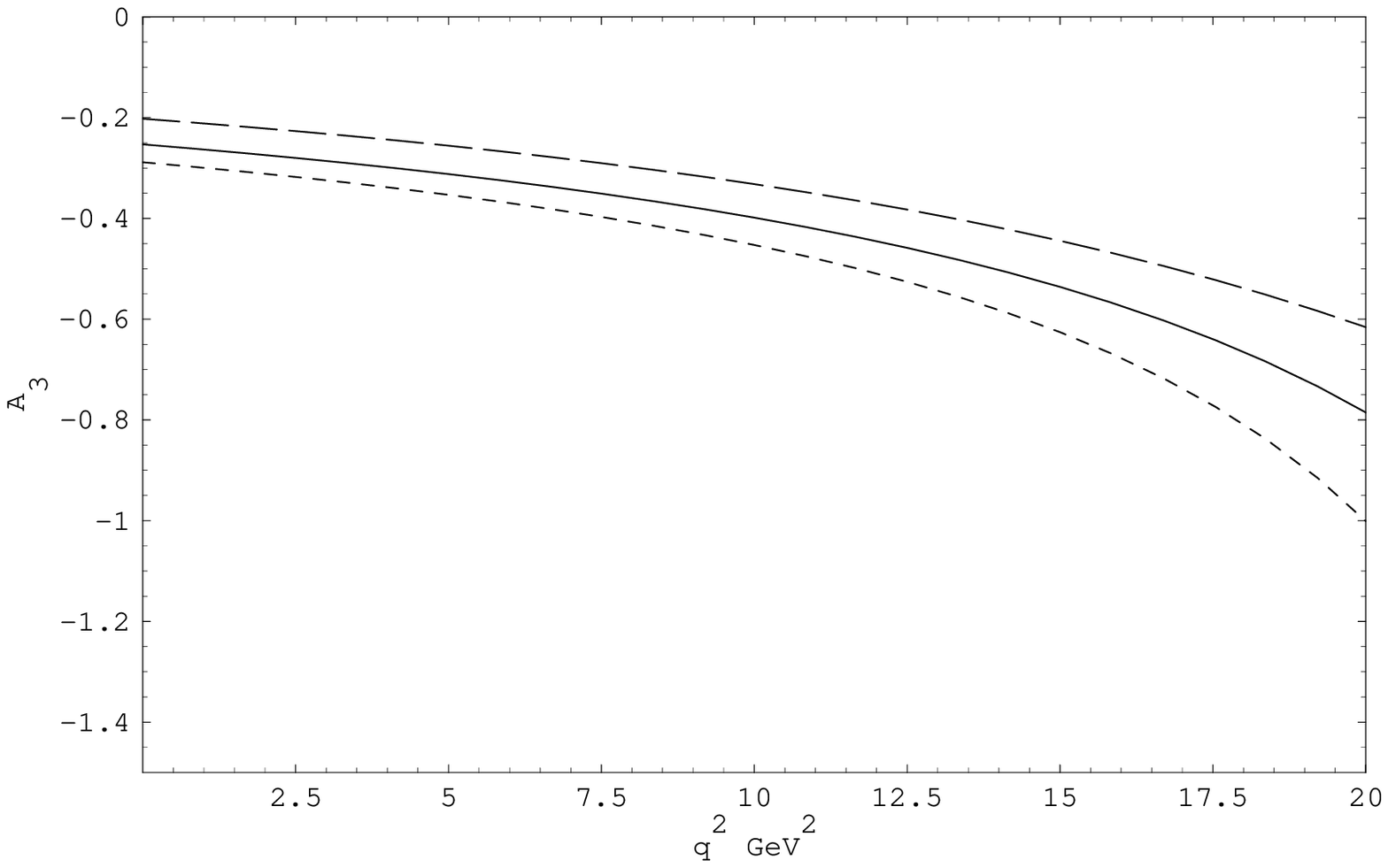}{Fig.2(c)}

\PICR{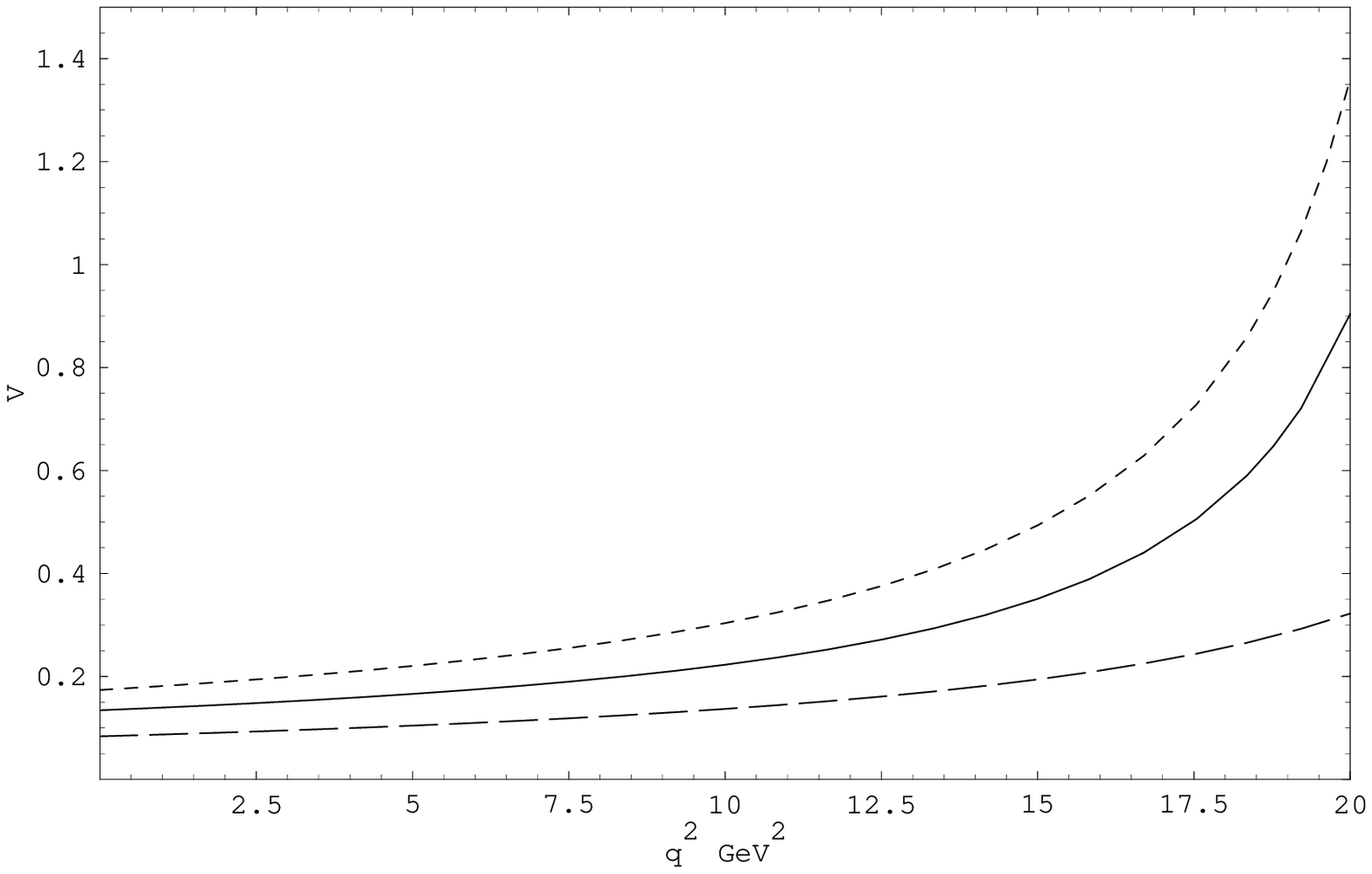}{Fig.2(d)}

\vspace{-2cm}
\centerline{
\parbox{12cm}{
\small
\baselineskip=1.0pt
Fig.2(a-d). Form factors obtained at the fixed Borel parameter $T=2.0$ GeV.  
The dashed, solid and dotted curves correspond to $s_0=$1.5, 2.1 and 2.7 GeV respectively.
} }

\end{document}